# Data-driven optimization of sparse sensor placement in thermal hydraulic experiments

Xicheng Wang[1*], Yun. Feng[2,3], Dmitry Grishchenko[1], Pavel Kudinov[1*], Ruifeng Tian[2,3], Sichao Tan[2,3]

[1]Division of Nuclear Science and Engineering, Royal Institute of Technology (KTH), Stockholm, Sweden.
[2]College of Nuclear Science and Technology, Harbin Engineering University, Harbin, China.
[3]Heilongjiang Provincial Key Laboratory of Nuclear Power System & Equipment, Harbin Engineering University, Harbin, China.

[*]Corresponding author: xicheng@kth.se, pkudinov@kth.se.

## ABSTRACT

Thermal-Hydraulic (TH) experiments provide valuable insight into the physics of heat and mass transfer and qualified data for code development, calibration and validation. However, measurements are typically collected from sparsely distributed sensors, offering limited coverage over the domain of interest and phenomena of interest. Determination of the spatial configuration of these sensors is crucial and challenging during the pre-test design stage. This paper develops a data-driven framework for optimizing sensor placement in TH experiments, including (i) a sensitivity analysis to construct datasets, (ii) Proper Orthogonal Decomposition (POD) for dimensionality reduction, and (iii) QR factorization with column pivoting to determine optimal sensor configuration under spatial constraints. The framework is demonstrated on a test conducted in the TALL-3D Lead-bismuth eutectic (LBE) loop. In this case, the utilization of optical techniques, such as Particle Image Velocimetry (PIV), are impractical. Thereby the quantification of momentum and energy transport relies heavily on readings from Thermocouples (TCs). The test section was previously instrumented with many TCs determined through a manual process combining simulation results with expert judgement. The proposed framework provides a systematic and automated approach for sensor placement. The resulting TCs exhibit high sensitivity to the variation of uncertain input parameters and enable accurate full field reconstruction while maintaining robustness against measurement noise.

# 1. INTRODUCTION

The design and safe operation of nuclear power systems depend on the ability to accurately estimate key thermal-hydraulic states to assess system performance during normal operation and accidental scenarios. These estimations are typically obtained using System Thermal Hydraulics (STH) codes (e.g. RELAP5 [1], GOTHIC [2]) and Computational Fluid Dynamics (CFD) codes (e.g. ANSYS Fluent [3]), which solve a set of parameterized Partial or Ordinary Differential Equations (PDEs or ODEs). However, the predictive capability of these codes is often limited by the epistemic uncertainties arising from incomplete knowledge.

To mitigate these uncertainties and improve model fidelity, dedicated Thermal-Hydraulic (TH) experiments are conducted at various scales. These experiments provide not only essential insights into the underlying heat and mass transfer phenomena but also qualified datasets for model development, calibration and validation. Nonetheless, measurements in large-scale facilities are often sparsely collected by sensors, providing limited coverage of the entire domain of interest and being susceptible to both systematic and random noise. For example, the PANDA facility, which is designed for containment safety research, includes vessels spanning over 10 m in height and 4 m in diameter. While, it is instrumented with only several dozen to a few hundred sensors such as Thermocouples (TCs), and pressure transducers [4][5]. Moreover, sensor placement is often restricted by extreme operating conditions (e.g. high temperatures, limited access ports [6]), and intrusive effects of instrumentation, making certain locations infeasible.

Sparse sensor configurations are also essential when direct measurement techniques are impractical. For instance, the optical based technique of Particle Image Velocimetry (PIV) has been broadly applied to record the velocity fields by tracking the illuminated particles. However, its applicability is significantly limited in multiphase or thermally stratified flows. In PANDA experiments involving direct contact condensation, PIV was unable to capture the main flow characteristics after steam condensation at high flow rates due to rapid bubble collapse and sharp temperature gradient [7][8]. As an alternative, the PPOOLEX facility deployed an array of 42 TCs in the vicinity of the injection orifice to monitor thermal field [9]. Moreover, PIV is entirely unsuitable for optically opaque working fluids, such as liquid metals, which are widely adopted in Gen IV reactor concepts. In such cases, model validation must rely on indirect measurements obtained from sensors like TCs [10].

Despite the increasing improvements in numerical algorithms and the development of STH and CFD codes, the validation of these codes for integral effect phenomena—system level responses emerging from coupled interactions among separate processes—remains a significant challenge [11][12]. Modeling such phenomena in integral test facilities typically involves a large number of Uncertain Input Parameters (UIPs) which may originate from insufficient knowledge of physical models, geometric configurations material properties, initial and boundary conditions. These UIPs are either not directly measurable or have limited accuracy in the experimental measurements.

Sensitivity Analysis (SA) is commonly employed to address these uncertainties by identifying the most influential UIPs on key model outputs, known as System Response Quantifies (SRQs). SA facilitates the UIPs calibration by prioritizing the most influential ones while the effect of remaining uncertainties is addressed through Uncertainty Analysis (UA) [11]. Calibration of UIPs is performed by evaluating the agreement between simulations and experimental measurements of SRQs. The SRQs should be defined to capture the most important physical phenomena relevant to the intended code application and should exhibit high sensitivity to the code UIPs. For instance, in the TALL-3D facility—a Lead-Bismuth Eutectic (LBE) loop consisting of three interconnected loops and a 3D test section that enables mutual feedback between natural circulation and complex 3D phenomena of thermal stratification and mixing— the selected SRQs for 3D section include: (i) LBE outlet temperature, (ii) temperature profiles along radial and vertical directions, (iii) the timing of peak temperatures during transients and (iv) the pressure drop across the test section [11].

The SRQs are measured by sensors, which are sparsely positioned over the entire domain of the test section. Pre-test analysis for instrumentation, including the definition of SRQ categories (e.g. temperature, pressure) and the spatial placement of corresponding sensors, is a non-trivial task. These decisions are often made based on experts' judgement, which may not guarantee that the selected SRQs are sufficiently sensitive to the UIPs. As the SRQs are directly informed by sensor readings, the spatial placement of the sensors plays a decisive role in determining the utility of the collected data. Therefore, optimizing sensor placement is critical to ensure that the measurements by these limited sensors exhibit highest sensitivity to variations in the UIPs

Optimal sensor placement generally seeks to maximize an objective, such as information criterial or reconstruction accuracy [13][14], by selecting configurations from a set of feasible options, often formulating the problem as a submodular selection task [15]. This problem can be efficiently addressed using greedy submodular techniques [15] or convex optimization [14] for sensor locations ranging from hundreds to thousands. However, engineering systems involve millions of grid points, making traditional optimization methods computationally impractical. Fortunately, the full fields of interest normally exhibit low-dimensional features, providing the applicability of Reduced-Order Models (ROMs) for dimension reduction, e.g. Proper Orthogonal Decomposition (POD) [16]. These approaches compress high-dimensional data by its low-dimensional representation with significantly fewer degrees of freedom, while preserving its essential characteristics. Recent study has demonstrated the use of empirical ROM interpolation techniques to identify optimal sensor locations with the goal of minimizing reconstruction errors [17]. Optimization incorporating spatial constraints is developed in [18].

This study develops a data-driven framework for optimal sensor placement in thermal-hydraulic experiments, with the specific goal of improving the utility of the experimental measurements for calibration of UIPs and code validation. The framework is demonstrated on a LBE experiment performed in the TALL-3D facility [10]. While Ultrasound Doppler Velocimetry (UDV) was initially intended to measure flow velocities within the 3D test section, its implementation was unsuccessful. Consequently, a dense array of TCs was installed on both the wall and internal regions of the test section to capture its thermal behavior. The TCs layout had been manually designed through an iterative process combining CFD simulation results with expert judgement. However, this approach was time-consuming and provided no guarantee of optimality.

The paper is organized as follows: Section 2 presents the methodology; Section 3 describes the dataset generated via SA using CFD, Section 4 discusses the results; and Section 5 concludes the work.

## 2. METHODOLOGY

The framework for the data-driven optimal sparse sensor placement is illustrated in Figure 1. It incorporates a sensitivity analysis to construct representative datasets, which are subsequently used in an optimization procedure to determine the optimal spatial configuration of sparse sensors. The detailed methodology is presented in the following subsections.

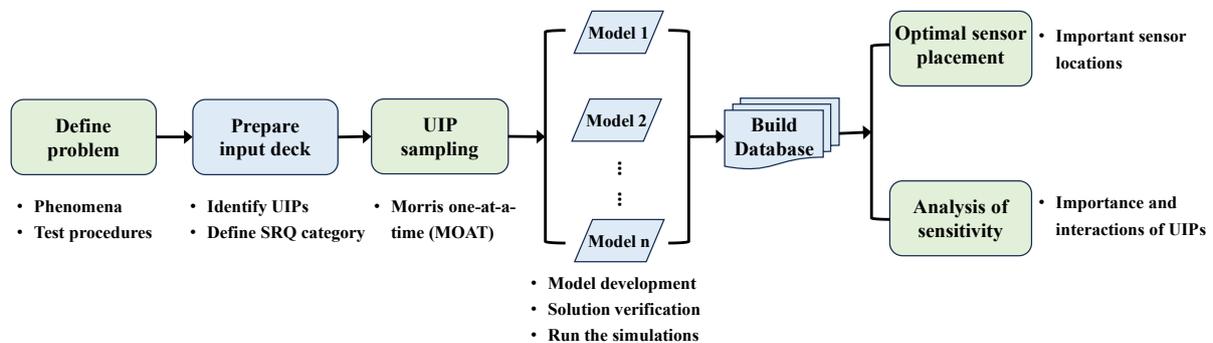

Figure 1. The framework for data-driven optimization of sparse sensor placement.

## 2.1. Sensitivity analysis to construct dataset

An original contribution of this work is the construction of a comprehensive dataset through global sensitivity analysis, to facilitate the data-driven optimization of sensor placement. This process consists of five main steps: (i) problem definition, (ii) input deck preparation, (iii) uncertainty sampling, (iv) simulation execution and (v) data collection.

Problem definition outlines the types of phenomena to be investigated and the procedural steps of the target test. The input deck includes all necessary parameters for carrying out pre-test simulations. A set of influential UIPs is identified based on their potential impact on the phenomena of interest. The UIPs and their respective ranges are initially determined through expert judgement and may be iteratively refined based on outcomes of the sensitivity analysis. Additionally, the categories of SRQ are specified that characterize system-level behavior which serve as the outputs for evaluating the sensitivity and performance of sensor configurations.

Subsequently, a global sensitivity analysis is performed using the Morris one-at-a-time (MOAT) method [24]. This approach enables efficient screening of UIPs by perturbing each parameter individually across a discretized input space and computing its effects on the SRQs. For each sampled input set, simulations are performed using STH or CFD codes. Prior to it, solution verification is needed to determine the proper spatial and temporal discretization. Finally, for each simulation case, multiple SRQ categories (e.g. temperature, velocity, pressure etc.) are extracted and organized into separate matrices, with each matrix corresponding to a distinct SRQ category and capturing its spatiotemporal evolution across all cases, as illustrated in Figure 2. These results form the basis for subsequent optimization of sparse sensor placement and sensitivity analysis.

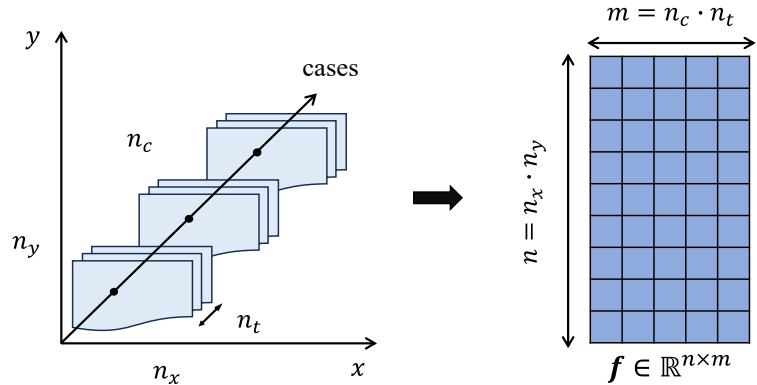

Figure 2. The collection of simulation results for a single SRQ category. Here $n_x$ and $n_y$ denotes the degrees of freedom in the spatial domain for the 2D problem, while $n_t$ and $n_c$ represent the number of time-step per simulation and the total number of simulated cases, respectively.

## 2.2. Sparse measurement and reconstruction

Sparse measurement on the entire domain (high-dimensional space) of a single SRQ category can be described by Eq. (1) and shown in Figure 3.

$$\boldsymbol{s} = \boldsymbol{Cf} \qquad (1)$$

where $\boldsymbol{f} \in \mathbb{R}^{n \times m}$ represents the collection of simulation results for a single SRQ category (Figure 2). $n$ is spatial dimension of the system and $m$ is the total number of snapshots. For steady-state simulations, $m$ equals to the number of simulated cases while for transient analysis, $m$ is the product of the temporal snapshots of each simulation and the number of simulations. $\boldsymbol{S} \in \mathbb{R}^{p \times m}$ denotes a SRQ measured by $p$ sensors from the entire domain of interest, e.g. temperature profiles along vertical direction in the test section. $C \in \mathbb{R}^{p \times n}$ represents the measurement selection operator consisting of $p$ sensors that are sparsely placed across the entire domain of interest ($p \ll n$) and it can be structured as:

$$C(\gamma) = \left[e_{\gamma_1}, e_{\gamma_2} \ldots e_{\gamma_p}\right]^T \tag{2}$$

where $e_{\gamma_i} \in \mathbb{R}^n$ denotes the canonical basis vectors with a unit entry at index $\gamma_i$ (sensor location) and zeros elsewhere.

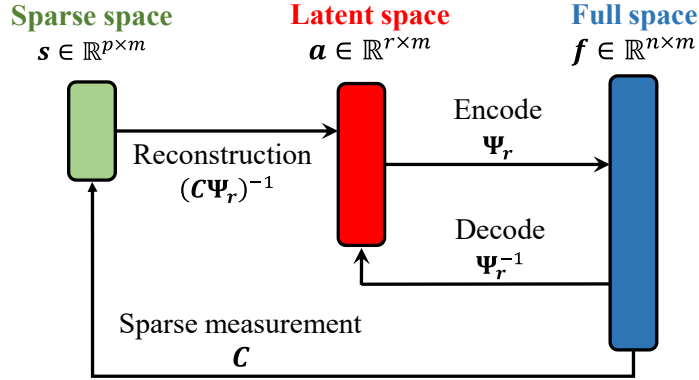

Figure 3. Sparse measurement from full (high-dimensional) space and reconstruction from sparse space.

The high-dimensional space $f$, despite complex spatial-temporal dynamics, normally exhibits low dimensional features, making them suitable for dimensionality reduction techniques. The primary objective of these techniques is to extract spatial modes that characterize the system, enabling a sparse representation in which the high-dimensional space is mapped to a lower dimensional latent space. In this work, we applied POD to represent $f$ in a low-rank form as:

$$f \approx \Psi_r a \tag{3}$$

where $\Psi_r \in \mathbb{R}^{n \times r}$ is spatial mode matrix where ($r \ll n$) and $a \in \mathbb{R}^{r \times m}$ is coefficient matrix indicating which few modes of $\Psi_r$ are active. These two matrices are solved by data-driven approach using the Singular Value Decomposition (SVD) by Eq. (4) where the leading $r$ left singular vectors consist of the desired POD modes $\Psi_r = u^*_{n \times r} \Sigma^*_{r \times r}$ (Figure 4).

$$f = u_{n \times n} \Sigma_{n \times m} v^T_{m \times m} \approx u^*_{n \times r} \Sigma^*_{r \times r} v^{T*}_{r \times m} \tag{4}$$

where '*' denotes the submatrices in grey are ignored. The matrix $f$ therefore can be approximated by the matrices with lower dimensions ($r \ll n$). The singular values (diagonal entries of $\Sigma$) represent the decreasing energy contributions of each subsequent mode and determine the truncation rank. In most fluid dynamic systems, the number of relevant degrees of freedom is much smaller than the overall data dimension, allowing for a significantly reduced choice of $r$.

Given this low-rank representation, the high-dimensional dataset $f$ can be directly reconstructed from the sparse measurement via the maximum likelihood estimate of the coefficient matrix by Eq. (5) which is also known as gappy POD [19].

$$\hat{f} = \Psi_r \hat{a} = \begin{cases} \Psi_r (C\Psi_r)^{-1} s, & p = r \\ \Psi_r (C\Psi_r)^{\dagger} s, & p > r \end{cases} \tag{5}$$

where $\hat{f}$ and $\hat{a}$ are reconstructed high-dimensional dataset $f$ and estimated coefficient matrix, respectively. $\dagger$ represents the Moore-Penrose pseudoinverse. The estimation remains well-posed if the number of sensors is at least equal to the dimension of the reduced basis (or latent space) i.e. $p \geq r$. Given the inherent compressibility of the TH simulations, i.e. high-dimensional field can be mapped to a reduced dimensional latent space, reconstructing $\hat{f}$ is feasible through strategically placed sensors with limited numbers.

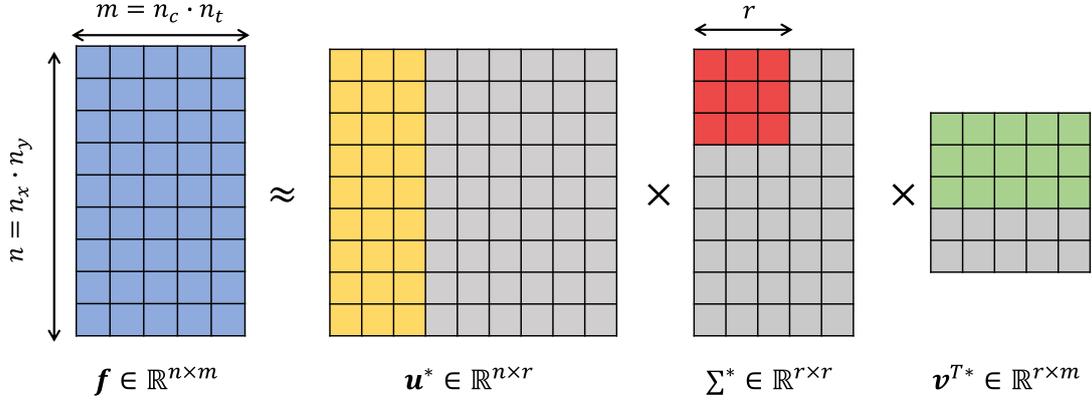

Figure 4. Schematic of dimensionality reduction using POD.

## 2.3. Optimal sensor placement

With the category of SRQ defined, the subsequent objective is to determine the optimal sensor locations that measurements by these sensors exhibit highest sensitivity to variations in the UIPs. Given the dataset $f$ is constructed by collecting SA results with varying UIPs (Figure 2), the optimization problem can be reformulated as selecting a set of sensors whose measurements enable accurate reconstruction of the full state $\hat{f}$, or equivalently, the mode coefficients $\hat{a}$ as expressed in Eqs. (5).

### 2.3.1. QR factorization with column pivoting

The accuracy and robustness of the reconstruction of $\hat{a}$ depend critically on the numerical conditioning of the matrix $C\Psi_r$. A well-conditioned $C\Psi_r$ ensures stable inversion, thereby minimizing the amplification of noise measurement in both $\hat{a}$ and $\hat{f}$. Consequently, the optimal sensor placement problem reduces to selecting a subset $\gamma = \{\gamma_1, \gamma_2, \dots, \gamma_p\}$ of spatial indices (as defined in Eq. (2)) such that the inversion of $C(\gamma)\Psi_r$ is optimally conditioned. For brevity, we denote $\Theta_\gamma = C(\gamma)\Psi_r$. The condition number can be indirectly bounded by optimizing the spectral content of $\Theta_\gamma^T \Theta_\gamma$, using metrics such as its determinant, trace or spectral radius.

In this work, QR factorization with column pivoting [17] is employed to maximize the determinant of the $\Theta_\gamma^T \Theta_\gamma$:

$$\gamma_* = \underset{\gamma, |\gamma|=p}{\operatorname{argmax}} |det(\Theta_\gamma^T \Theta_\gamma)| = \underset{\gamma, |\gamma|=p}{\operatorname{argmax}} \prod_i |\sigma_i(\Theta_\gamma^T \Theta_\gamma)| \qquad (6)$$

where $p$ is the number of sensors and $\sigma_i(\Theta_\gamma^T \Theta_\gamma)$ is the $i$-th singular value of $\Theta_\gamma^T \Theta_\gamma$. When $p = r$ (i.e. number of sensors equals to number of extracted modes), Eq. (6) reduces to optimizing $det(\Theta_\gamma)$.

QR column pivoting, originally introduced by Businger and Golub in the 1960s for solving lease-squares problems [21], decomposes the matrix $X \in \mathbb{R}^{m \times n}$ as:

$$XP^T = QR \qquad (7)$$

where $Q \in \mathbb{R}^{m \times m}$ is an orthogonal matrix, $R \in \mathbb{R}^{m \times n}$ is an upper-triangular matrix, $P \in \mathbb{R}^{n \times n}$ is a permutation matrix encodes the column reordering determined by the pivoting process. During this process, the algorithm iteratively selects the column with the largest Euclidean norm (e.g. two-norm) as the next pivot. Subsequently, the orthogonal projection of this pivot column is subtracted from all remaining columns (see Algorithm 1). The orthogonal projection is achieved using a Householder reflector $H$, which reflects a vector across a hyperplane designed to zero out selected components. (see Algorithm 2). Specifically, at $k$-th iteration, the Householder reflector is constructed as:

$$u = \frac{v + \alpha e_1}{\|v + \alpha e_1\|_2}, \quad \alpha = sign(v_1)\|v\|_2 \tag{8}$$

$$H = I - 2uu^T \tag{9}$$

where $v$ is the sub-vector of the $k$-th column of $R$, starting from row $k$ to $m$, $v_1$ is the first component of $v$, $e_1$ is the first canonical basis vector of the corresponding dimension, $sign(\cdot)$ returns the sign of its argument, and $I$ is the identify matrix.

The iterative pivoting strategy provides an approximate greedy approach to the optimization problem in Eq. (6) [17]. The method effectively enlarges the volume of the selected submatrix by enforcing a diagonal dominance pattern [20], as expressed in Eq. (10). Since the absolute value of a matrix determinant—interpreted as its volume—can be expressed as the product of the diagonal elements of $R$ (Eq. (11)), the pivoting implicitly maximizes the determinant.

$$\sigma_i^2 = |r_{ii}|^2 \geq \sum_{j=i}^{k} |r_{jk}|^2, 1 \leq i \leq k \leq m \tag{10}$$

$$|det\ X| = \prod_i \sigma_i = \prod_i |r_{ii}| \tag{11}$$

Recall that $\Psi_r \in \mathbb{R}^{n \times r}$ where each column corresponds to the spatial pattern of a POD model and each row represents a candidate sensor location. Applying QR factorization with column pivoting to $\Psi_r^T$, yields $r$ pivot columns that optimally sample the $r$ basis modes, expressed as $\Psi_r^T P^T = QR$. The permutation matrix $P$ serves as the measurement selection operator $C$, encoding the indices of the selected sensors. In the oversampled case, where the number of sensors $p$ exceeds $r$, the optimization can be performed via the QR factorization of $(\Psi_r \Psi_r^T) P^T = QR$ [17].

### 2.3.2. Spatial constraints

In this work, we consider two types of spatial constraints [18], as illustrated in Figure 5:

1) **Region constraints**: This type of constraint arises when certain regions are unavailable for sensor placement. In practical application, sensor locations are often constrained by factors such as extreme operating conditions or the intrusive effects of instrumentation. The former case can render certain locations entirely infeasible whereas intrusive effects occur when the presence of sensors perturb the local physical system, potentially influencing the accuracy and representativeness of the acquired measurements.
2) **Distance constraints**: This type of constraint enforces a minimum distance between selected sensors. During the iterative pivoting process, the set of allowable locations is adaptively updated to exclude regions that fall within the prescribed distance from selected sensors.

The indices corresponding to constrained regions are identified and the column-wise norms of these locations are set to zero, ensuring that they are not considered in the iterative pivoting procedure (section 2.3.1). The entire process is summarized in Algorithm 3.

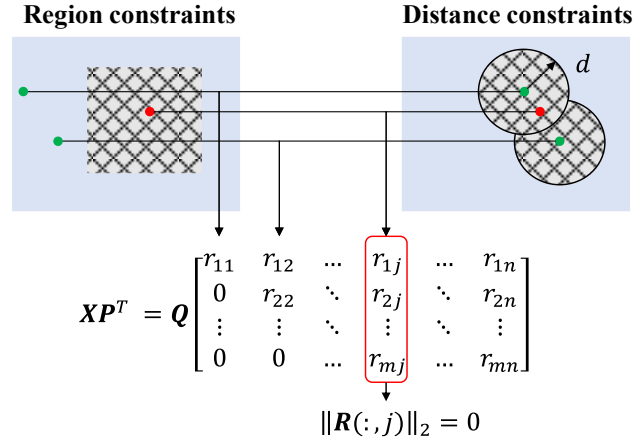

Figure 5. Greedy selection of the next pivot column (i.e. sensor location) under region and distance constraints, with constrained locations shaded in grey. Column-wise norms in the constrained regions are set to zero.

---

Algorithm 1: QR factorization with column pivoting

---

**Inputs**: $X$ = input matrix
**Output**: $Q$ = orthogonal matrix, $R$ = upper-triangular matrix, $P$ = permutation matrix
**Procedure**: $[Q, R, P] = qr\_pivot(X)$

1: $m, n \leftarrow size(X)$
2: $R \leftarrow X$
3: $P \leftarrow eye(n), \ Q \leftarrow eye(m)$      % identity matrix
4: **for** $k \leftarrow 1$ to $m$ **do**:
5:     **dlens** = $vecnorm(R(k:m, k:n), 2, 1)$      % vector-wise 2-norm along columns
6:     $l \leftarrow argmax$ **dlens**      % index of maximum dlens
7:     $idx1 \leftarrow k, \ idx2 \leftarrow k - 1 + l$
8:     Swap: $R(:, [idx1, idx2]) \leftarrow R(:, [idx2, idx1])$      % update $R$
9:     Swap: $P(:, [idx1, idx2]) \leftarrow P(:, [idx2, idx1])$      % update $P$
10:    $[Q, R] \leftarrow householder(Q, R, k)$      % household projection
11: **end**
**return** $Q, R, P$

---

---

Algorithm 2: Householder projection

---

**Inputs** : $Q$, $R$ = orthogonal and upper-triangular matrices, $k$ = k-th iteration
**Output**: $Q$, $R$ = matrices after householder transformation
**Procedure:** $[Q, R] = householder(Q, R, k)$

1: $m, n \leftarrow size(R)$
2: $v \leftarrow R(k{:}m, k)$
3: $e_1 \leftarrow zeros(size(v))$
4: $e_1(1) \leftarrow 1$
5: $\alpha \leftarrow sign(v(1))\|v\|_2$
6: $u \leftarrow (v + \alpha e_1)/\|v + \alpha e_1\|_2$
7: $H \leftarrow eye(size(uu^T)) - 2uu^T$
8: $R(k{:}m, k{:}n) \leftarrow H \times R(k{:}m, k{:}n)$
9: $Q(:, k{:}m) \leftarrow Q(:, k{:}m) - Q(:, k{:}m) \times 2uu^T$
**return** $Q, R$

---

Algorithm 3: Optimal sensor placement with spatial constraints

---

**Inputs**: $\Psi_r$ = POD basis matrix, *constraint_fun* = handle of constraint function, $d$ = minimum distance between sensors, $p$ = number of sensors
**Output**: $\gamma$ = sensor indices
**Procedure:** $[\gamma] = qr\_pivot\_constraints(\Psi_r, constraint\_fun, d, p)$

1: $m, n \leftarrow size(\Psi_r^T)$
2: **if** $p > m$ **then**                                    % oversample
3:     $R \leftarrow \Psi_r \Psi_r^T$
4: **else**
5:     $R \leftarrow \Psi_r^T$
6: **end**
8: $\gamma \leftarrow [1, 2, \cdots, n]$
9: **constraints_idx** $\leftarrow zeros(size(\gamma))$
10: **for** $k \leftarrow$ 1 to $p$ **do**:
11:     dlens = $vecnorm(R(k{:}m, k{:}n), 2, 1)$           % vector-wise 2-norm along columns
12:     % user defined function for region and distance constraints
13:     **constraints_idx** $\leftarrow constraint\_fun(d)$
14:     **dlens**(*constraints_idx*) = 0                    % exclude constrained locations
15:     $l \leftarrow argmax$ **dlens**                     % index of maximum dlens
16:     $idx1 \leftarrow k$, $idx2 \leftarrow k - 1 + l$
17:     Swap: $R(:, [idx1, idx2]) \leftarrow R(:, [idx2, idx1])$    % update $R$
18:     Swap: $\gamma([idx1, idx2]) \leftarrow \gamma([idx2, idx1])$    % update $\gamma$
19:     $[\sim, R] \leftarrow householder(\sim, R, k)$      % household projection
20: **end**
**return** $\gamma(1{:}p)$

---

## 2.4. Demonstration on parameterized gaussian distribution

The proposed framework is first demonstrated using a parameterized Gaussian distribution $\mathcal{N}(\mu, \sigma^2)$ with $x = [-10: 0.01: 10]$, $\mu = [-2, 3]$, $\sigma = [0.5: 0.2: 6.5]$ as illustrated in Figure 6a. POD is applied to the full state dataset $f \in \mathbb{R}^{1000 \times 62}$, where 56 cases are used for training and the remaining 6 cases are used for testing. The POD analysis shows that 5 dominant modes are sufficient to capture 99.0% of the dataset's variability (a.k.a. cumulative energy contribution). Optimal sensor placement on the POD basis matrix $\Psi_r^T$, subject to a minimum sensor spacing of 0.25, results in 5 optimal sensors located as shown in Figure 6a. These optimized sensors coincide with regions of high variability, with the two most important sensors at $x = -2$ and $x = 3$.

The original profile can be reconstructed through readings of these sensors using Eq. (5). The reconstruction performance under noisy measurements is compared across different sensor configurations in Figure 6b. Sensor noise is simulated as a Gaussian random variable $\mathcal{N}(0, \sigma_{noise}^2)$. For each sensor configuration and noise level $\sigma_{noise}$, the reconstruction is repeated 10 times to ensure sufficient statistics, and the normalized mean squared error (NMSE) is averaged over it. The results show that randomly placed fail to estimate the original profile, whereas the optimally placed sensors achieve significantly better performance. The reconstruction accuracy decreases with increasing noise level while this effect can be mitigated by increasing the number of sensors.

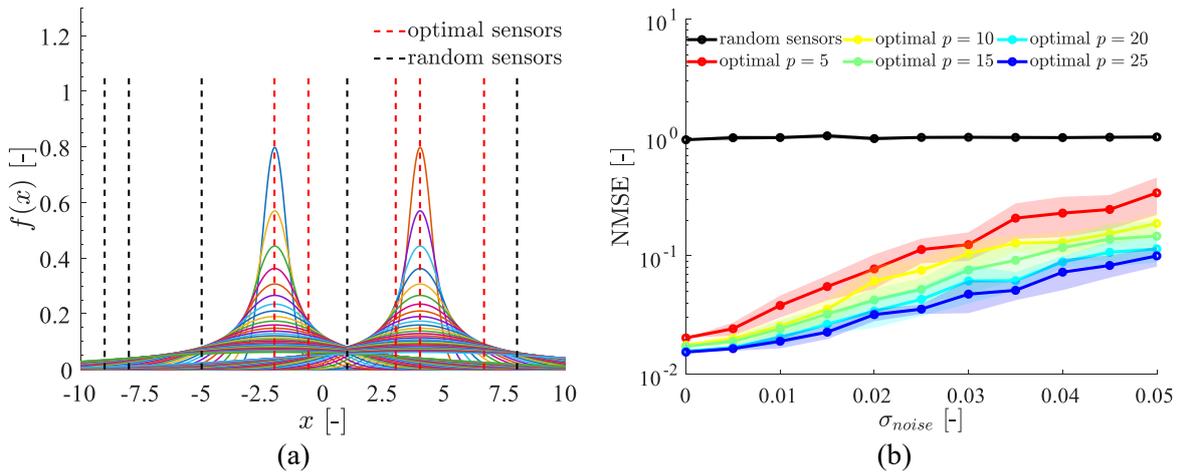

Figure 6. (a) Optimal sensor placement for a parameterized gaussian distribution, illustrating the locations of 5 optimally placed sensors in comparison to randomly placed sensors. (b) NMSE of full-state reconstruction from noisy sensor measurements, evaluated for different numbers of sensors. Each NMSE corresponds to the mean over 10 independent reconstructions at a given noise level, with the shaded area indicating the $\pm 1\sigma$ standard deviation.

## 3. DATASET CONSTRUCTION

In this section, we describe the detailed steps of constructing the dataset by performing a systemic sensitivity analysis using CFD code for an LBE experiment performed in TALL-3D facility.

### 3.1. TALL-3D facility

TALL-3D is an LBE loop-type facility built and operated at the Royal Institute of Technology (KTH) in Stockholm, Sweden. It is designed to provide measurements for development and validation of coupled STH and CFD codes. As shown in Figure 7, the facility consists of three vertical legs: (i) the Main Heater (MH) leg, (ii) the Test Section (TS) leg featuring a 3D pool-type section, and (iii) the Heat Exchange (HX) leg. The MH leg contains an electric heater for LBE heating, while the lower part of the HX leg houses an electric permanent-magnet pump that can operate in forced-circulation mode or remain deactivated to allow natural circulation. The HX leg is equipped with a countercurrent, double-pipe HX to remove heat from primary loop to the secondary oil loop. Further details of the loop design can be found in [10].

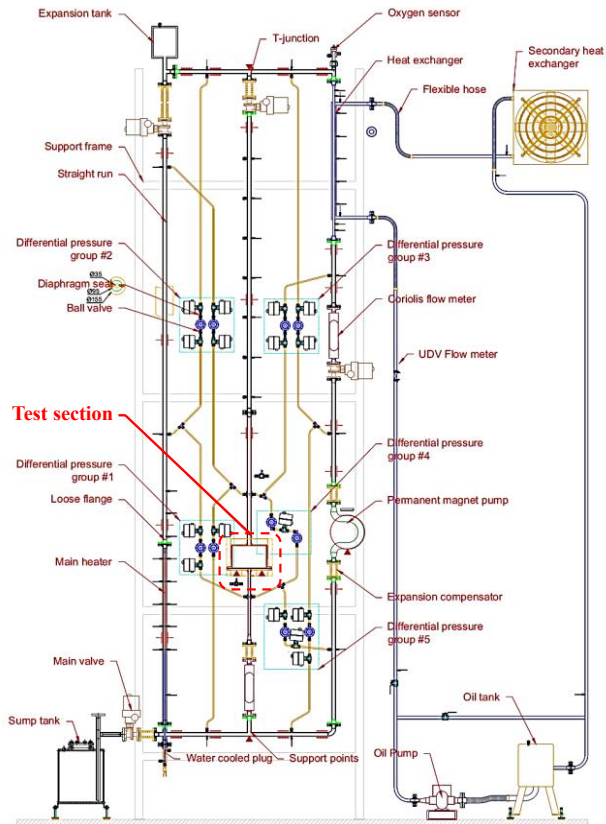

Figure 7. Schematic of the TALL-3D facility.

The pool type test section (Figure 8a) is designed to facilitate interactions between natural circulation and complex 3D phenomena such as thermal stratification and mixing. It contains an axisymmetric cylindrical vessel made by stainless steel with an inlet pipe positioned at the bottom and an outlet pipe at the top. A 15 kW band heater is mounted circumferentially around the upper region of the vessel to promote thermal stratification. A Circular Inner Plate (CIP) is installed within the test section to redirect flow to enhance pool mixing. The section allows the measurement of mass flow rate, outlet temperatures and pressure drop across the section.

The test section is instrumented with a large number of TCs to provide sufficient measurements of the thermal behavior of the pool. The sensor layout was determined through an iterative process that integrated CFD simulation results with expert judgement, aiming to ensure adequate coverage of key thermal phenomena such as thermal stratification, mixing, and jet impinging on the circulate plate. However, this manual design process was time-consuming and there was no guarantee that the resulting layout was indeed optimal in terms of maximizing measurement sensitivity to variations in UIPs. In this work, the data-driven framework introduced in Section 2 is applied to re-design the TCs instrumentation to improve design efficiency and utility of the measurements.

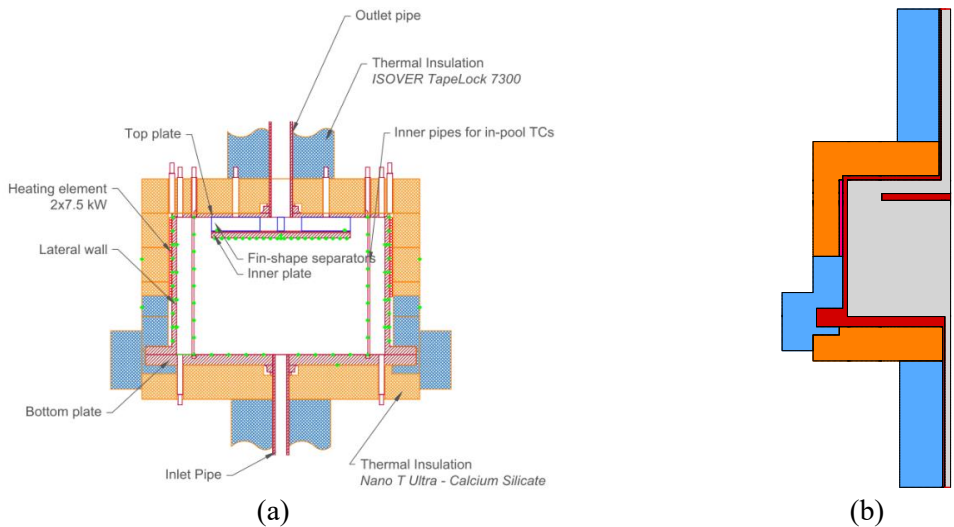

(a) (b)

Figure 8. Schematic of TALL-3D (a) test section [10] and (b) its corresponding CFD mesh. Green dots indicate the locations of thermocouples.

### 3.2. Reference case

The boundary conditions for the sensitivity analysis are derived from measurements recorded during the transient case TG03.S302: S3_FN_11_01 [22] as shown in Figure 9. This scenario represents a forced-to-natural circulation transient under constant power in the 3D test section. During the forced circulation phase, the LBE flow moves upward through the MH and TS legs, while flowing downward in the HX leg. After the deactivation of the pump (~800s), the flow rate in the TS leg decreases significantly, leading to a temporary flow reversal, as indicated by a negative flow rate. After a certain period, the flow direction returns to its original orientation and remains nearly constant.

Throughout the transient, a mutual interaction is observed between the loop flow dynamics and the stratification/mixing in the 3D test section. Accurate prediction of these coupled phenomena is essential for safety analysis. Therefore, this transient is selected as the reference case for the present study. The boundary conditions for the 3D test section, including LBE flow rate, inlet temperature and heater power, are provided for the subsequent CFD simulations.

It should be noted that obtaining accurate flow and temperature profiles, particularly the ones during the transient from forced to natural circulation, is often not feasible at the preliminary stage of experimental design, where prior knowledge is limited. Pre-test simulation, for instance using STH codes, can provide approximate estimates of the boundary conditions. Measurements from the subsequent tests can then be used to refine these inputs, enabling iterative updates of the sensor layout.

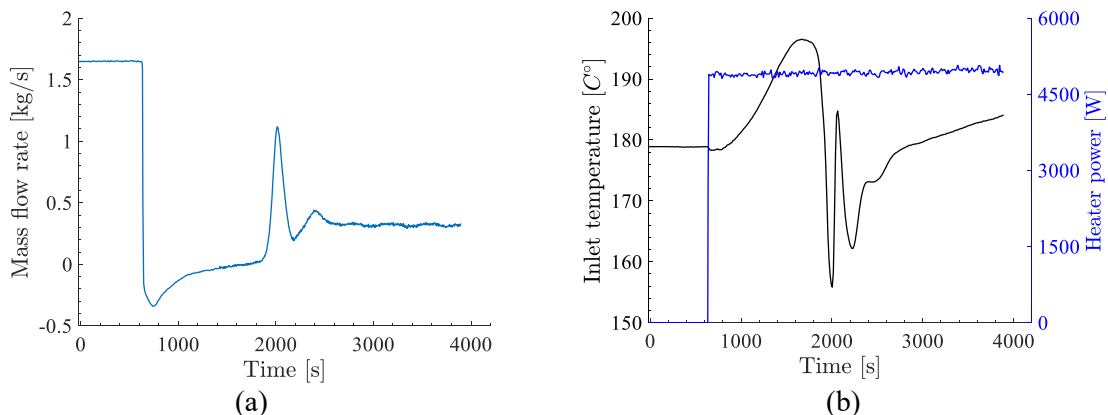

(a) (b)

Figure 9. LBE (a) mass flow rates, (b) inlet temperature and heater power of the 3D test section during TG03.S302: S3_FN_11_01.

## 3.3. Description of CFD model

Since the objective of this work is to determine optimal sensor placement within the 3D test section, our sensitivity analysis only focuses on this region. The computation domain and mesh of this section are illustrated in Figure 8. The domain consists of a LBE fluid region and four solid regions representing the steel vessel, heater and two different thermal insulation materials. 2D solver was applied to leverage the axis-symmetry of the test section and concerned phenomena, which is particularly useful in terms of sensitivity and uncertainty analysis. This axis-symmetry assumption was validated based on temperature profiles measured inside the test section. Mesh size was determined by mesh sensitivity analysis performed in [11].

Turbulent was solved using Realizable $k - \varepsilon$ model with buoyancy effects on turbulence production. Near wall profiles were resolved by standard wall functions. A constant turbulent Prandtl number 0.85 was employed to account for the effect of turbulence on the energy equation. Second-order upwind schemes were used for spatial discretization of momentum, energy, $k$ and $\varepsilon$ equations. Convergence criteria were set $1e-6$ for the energy equation, $1e-3$ for the continuity equation, and $1e-4$ for other variables. Segregated solver was used during the forced circulation stage and coupled solver was applied during the natural circulation stage [11].

LBE properties are summarized in Table 1, while the properties of other materials are detailed in [11]. Mass flow rate (Figure 9a) and pressure outlet were used as inlet and outlet boundary conditions. Heating during the natural circulation stage was modeled using a volumetric heat source with power history shown in Figure 9b. Heat loss to the environment was simulated using convective boundary condition with fixed ambient temperature and heat transfer coefficient on the external walls of the insulation. Thermal resistance due to non-ideal contact and the presence of the thin gasket was not considered.

Table 1. Material properties of LBE [23]

| Property | Correlation | Unit |
|---|---|---|
| Thermal conductivity | $3.284 + 1.617 \times 10^{-2} - 2.305 \times 10^{-6} \cdot T^2$ | $W/(m \cdot K)$ |
| Density | $11065 - 1.293 \cdot T$ | $kg/(m^3)$ |
| Specific heat capacity | $164.8 - 3.94 \times 10^{-2} \cdot T - 1.25 \times 10^{-5} \cdot T^2 - 4.56 \times 10^5 \cdot T^{-2}$ | $J/(kg \cdot K)$ |
| Dynamic viscosity | $4.94 \times 10^{-4} \cdot \exp(754.1/T)$ | $Pa \cdot s$ |

## 3.4. Sensitivity analysis

SA analysis was performed by MOAT method [24] (see Section 2.1), which is well-suited for a large complex system involving many UIPs. MOAT has been implemented within the Dakota (Design Analysis Kit for Optimization and Terascale Applications) toolkit [25]. A total of 19 UIPs were investigated, with 400 sample cases encompassing parameters from physics models (e.g. Figure 10a), material properties, and boundary conditions, as summarized in Table 2.

Each simulation ran for 3600s, generating 72 snapshots with 50s intervals. An example of the predicted LBE outlet temperature across the 400 cases is shown in Figure 10b. In this study, the selected category of SRQ is the temperature field. The complete collection of all temperature results forms a full state matrix $f$, with dimensions $49009 \times 28000$. This matrix was constructed using 28800 snapshots, each containing 49009 spatial nodes including the solid regions.

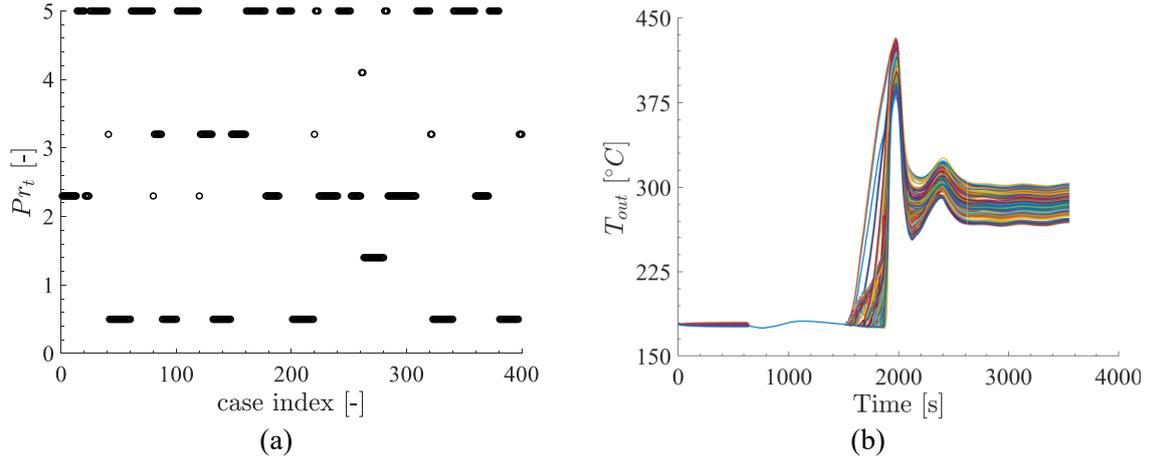

Figure 10. Sensitivity analysis of (a) sampling of turbulent Prandtl number and (b) predicted LBE outlet temperature during reference transient.

Table 2. Ranges for uncertain input parameters used in sensitivity analysis

| Category | | # | # Parameter | Ranges | Unit | Reference |
|---|---|---|---|---|---|---|
| Physics models | | 1 | HTC of insulation to air | $5 - 50$ | $W/(m^2 \cdot K)$ | Engineering toolbox |
| | | 2 | Turbulent Prandtl number | $0.5 - 5$ | - | |
| Materials | LBE | 3 | Density | $\pm 0.8\%$ | $kg/m^3$ | OECD LBE Handbook [23] |
| | | 4 | Dynamic viscosity | $\pm 8\%$ | $Pa \cdot s$ | |
| | | 5 | Thermal conductivity | $\pm 15\%$ | $W/(m \cdot K)$ | |
| | | 6 | Specific heat capacity | $\pm 7\%$ | $J/(kg \cdot K)$ | |
| | Isover wool | 7 | Density | $65 - 90$ | $kg/m^3$ | Manufacturer specifications |
| | | 8 | Thermal conductivity | $\pm 5.0\%$ | $W/(m \cdot K)$ | |
| | | 9 | Specific heat capacity | $800 - 900$ | $J/(kg \cdot K)$ | |
| | Nano T Ultra | 10 | Density | $65 - 90$ | $kg/m^3$ | |
| | | 11 | Thermal conductivity | $-5\% \sim +50\%$ | $W/(m \cdot K)$ | |
| | | 12 | Specific heat capacity | $\pm 10\%$ | $J/(kg \cdot K)$ | |
| | Strainless steel 316 L | 13 | Density | $230 \pm 5\%$ | $kg/m^3$ | |
| | | 14 | Thermal conductivity | $\pm 5\%$ | $W/(m \cdot K)$ | |
| | | 15 | Specific heat capacity | $\pm 5\%$ | $J/(kg \cdot K)$ | |
| Boundary conditions | | 16 | LBE mass flow rate | $\pm 0.02$ | $kg/s$ | |
| | | 17 | LBE inlet temperature | $\pm 2$ | $K$ | Calibrated using TALL-3D data |
| | | 18 | Ambient air temperature | $20 \sim 40$ | °C | TALL-3D data |
| | | 19 | Heater power | $\pm 5\%$ | $W$ | Manufacturer specifications |

## 4. RESULTS AND DISCUSSION

In this section, we applied the data-driven framework to the full state temperature field $f$, obtained from the sensitivity analysis to identify the optimal sensor placement under spatial constraints. The dataset used for training is the fluctuating fraction of the variables, defined as $f = f - \bar{f}$, where $\bar{f}$ denotes the matrix averaged over all snapshots. The dataset was partitioned into a training set (85%) to fit the model (i.e. Eqs.(3)-(5)) and a testing set (15%) to evaluate model performance.

## 4.1. Optimal sensor placement with spatial constraints

Applying POD (Eq. (4)) to the temperature field $f$ indicates that 24 POD modes (i.e. $\Psi_r \in \mathbb{R}^{49009 \times 24}$) can optimally represent 99.9% of the total variability (0.999 cumulative variance) of the full state temperature, as shown in Figure 10a. The first two modes account for the majority of the variability (93.6%). The first mode corresponds to a buoyancy-driven regime which is a typical flow condition during natural circulation. The second mode exhibits inertia-dominant behavior, representing the condition during forced circulation (Figure 10b).

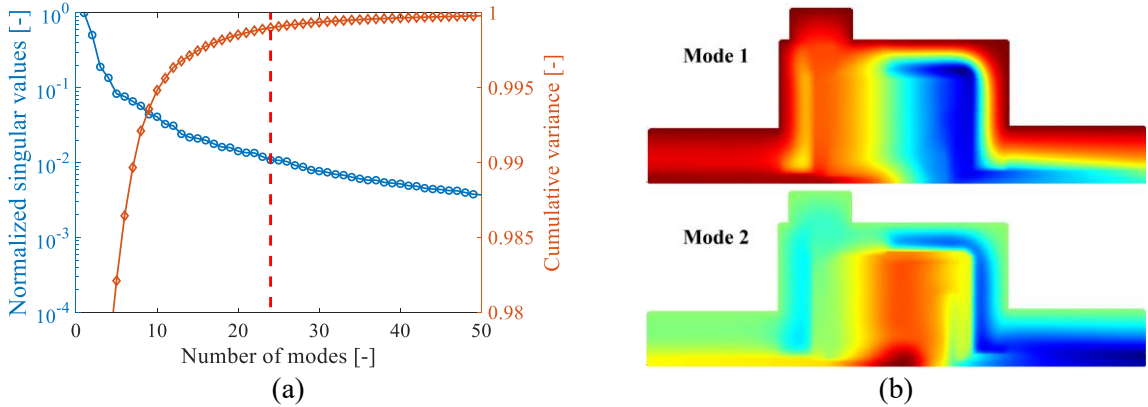

Figure 11. (a) Cumulative variance and normalized singular values as a function of the number of modes for temperature field and (b) corresponding first and second modes which account for 93.6% variability.

The optimization of sensor placement is performed using QR factorization with column pivoting applied to $\Psi_r^T$. Figure 12 shows the sensor locations and their relative importance under unconstrained conditions. To achieve sufficient accuracy in reconstructing the full temperature field, a minimum of 24 sensors (denoted as $p$) is required, corresponding to the number of extracted POD modes [17]. The unconstrainted placement indicates a cluster of sensors located along the central axis, with the remaining sensors distributed at the CIP and thermal insulation. The sensitivity of sensor readings to UIPs is assessed through case-wise variances calculated using Eqs. (12) and (13). A comparison of the variances for selected sensors is presented in Figure 13.

$$\sigma_T^2(t) = \frac{1}{N} \sum_{i=1}^{N} \left( T_i(t) - \bar{T}(t) \right)^2 \tag{12}$$

$$\bar{T}(t) = \frac{1}{N} \sum_{i=1}^{N} T_i(t) \tag{13}$$

where $N = 400$ is the number of simulation cases, $T_i(t)$ is the sensor reading at time-step $t$ for the $i$-th simulation case, and $\bar{T}(t)$ is the mean sensor reading across all $N$ cases.

It can be observed that the most influential sensors capture distinct thermal-hydraulic phenomena, with their peak variances occurring at different stages of the transient. For example, sensor #1, located at the corner of the external insulation, exhibits consistently high variability throughout the entire transient. This sensor directly measures the UIP associated with the ambient air temperature (Table 2). Sensor #2, positioned at the outlet, reflects the integral heat transfer across the test section. Its highest variability occurs between 1700~2200s, corresponding to the period when the inlet temperature oscillates significantly (Figure 9b). Sensor #3 captures the effect of the jet impinging on the CIP and reaches peak variability during the rapid change in inlet LBE mass flow rate, when it sharply increased to 1.1 kg/s before dropping to 0.2 kg/s. Sensor #6, located at the center of the T-shape channel above the CIP, shows its highest variability during the flow reversal transient, i.e. the mass flow rate becomes positive after pump deactivation. For comparison, a randomly placed sensor exhibits considerably lower variability, indicating a much weaker sensitivity to the variations in UIPs.

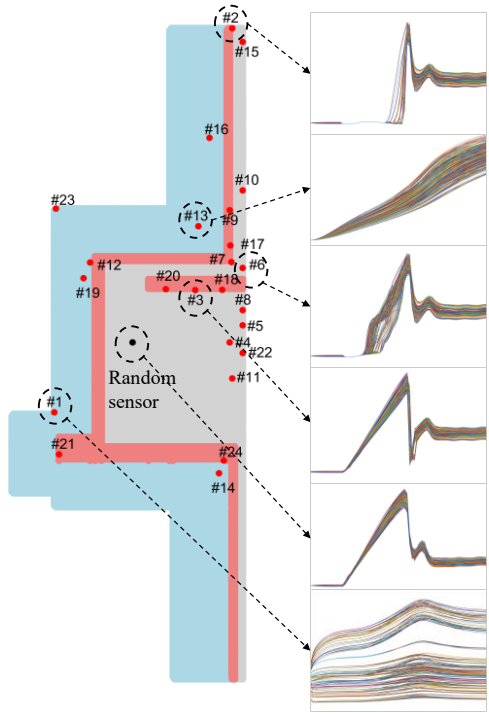

Figure 12: The ranking of unconstrained optimal placement of 24 sensors, along with selected sensor readings from 400 simulation results in the sensitivity study. The marker (●) indicates a randomly placed sensor that exhibits relatively low variation in readings across cases (see Figure 13).

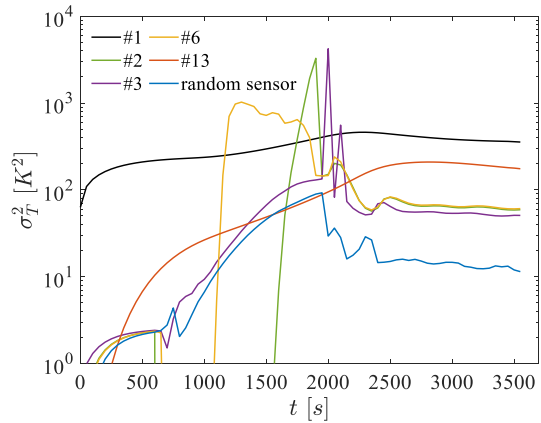

Figure 13: Variances of temperature profiles across the 400 sensitivity study results (calculated using Eqs. (12) and (13)) for the selected sensors shown Figure 12.

However, the unconstrained placement is limited by intrusive effects where the sensors clustered along the central axis would disturb the flow and disrupt the axisymmetric structure. This was the primary concern for the predetermined layout (green circles in Figure 14b). Another limitation is that sensors cannot be easily installed within the insulation or steel vessel. Accordingly, the constrained optimization method introduced in Section 2.3.2 is applied. The accessible regions include (i) the external surface of the insulation, (ii) the external and internal surfaces of the steel vessel, (iii) the outer LBE flow region, shown in grey in Figure 14a. In addition, a minimum spacing of 10 mm is imposed. The resulting constrained placement yields a configuration that closely matches the predetermined sensor locations used in previous TALL-3D experiments (Figure 14c). The ranking of sensor importance is presented in Figure 14d.

The key difference between the unconstrained and constrained layouts lies in the placement of sensors along the central axis of the test section. In the unconstrained condition, these sensors provide direct measurements of the LBE jet temperature profile, thereby capturing the dominant convection and diffusion processes. They also record the thermal stratification that develops within the test section during heater operation. In constrained configuration, however, convection and diffusion effects are instead monitored indirectly by additional sensors positioned at the CIP, while thermal stratification is captured by three sensors (#14, #18, and #22) placed along the inner surface of the sidewall in the vertical direction.

Furthermore, we also perform oversampled sensor placement (i.e. $p > r$, the number of sensors exceeding the number of extracted POD modes) by applying QR column pivoting to $\boldsymbol{\Psi}_r \boldsymbol{\Psi}_r^T$. In this case, 44 sensors are selected to match the number of predetermined TCs in TALL-3D (Figure 14e). For comparison, three additional layouts obtained by random placement are also considered and will be discussed in the next section (Figure 14e, f, g).

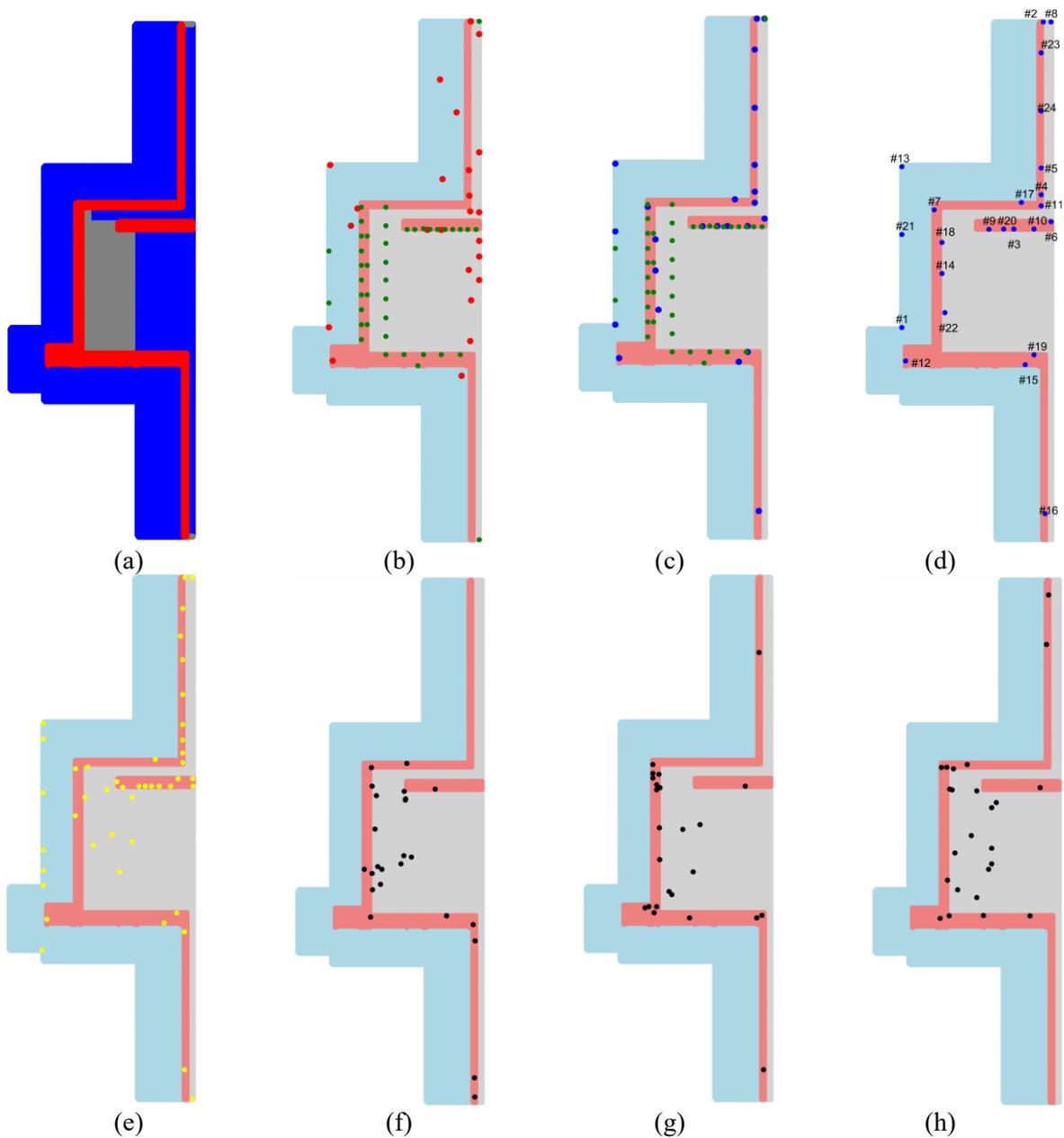

(a) (b) (c) (d)

(e) (f) (g) (h)

Figure 14. Illustration of sensor placement. (a) Constrained regions are shown in blue (excluding the external surface of the insulation), while accessible LBE and steel vessel regions are indicated in grey and red. Sensor locations are shown for (b) unconstrained optimal placement with $p = 24$ (●); (c) constrained optimal placement with $p = 24$ (●); and (e) constrained optimal placement with $p = 44$ (●). Predetermined sensor locations in TALL-3D with $p = 44$ (●) (Figure 8a) are displayed for comparison in (b)(c). The ranking of constrained optimal placement with $p = 24$ is illustrated in (d). Randomly placed sensors with $p = 24$ (●) are shown in (f), (g), and (h).

## 4.2. Reconstruction by sparse sensing

In this section, we evaluate the optimality of various sensor configurations by comparing their reconstruction errors, specifically the Normalized Mean Square Error (NMSE), on the testing dataset. The reconstruction process is achieved through Eq. (5) where $s$ represents sparse measurements obtained from TC readings.

Reconstruction performance is expected to degrade in the presence of noise. In this context, noise may refer to the uncertainty associated with thermocouple measurements. Measurement noise normally consists of a ~0.1K inherent fluctuation caused by the oscillation of the electrical signal and a fixed offset varying from 0.1 to 2K during the test. This offset can be reduced by performing the calibration test. Moreover, discrepancies in sensor positioning during experiments may also introduce noise.

To assess the robustness of the designs, random noise was added to the sparse temperature inputs. The noise was modeled as a normal distributed variable, $N(0, \sigma_{noise}^2)$. A total of 7 configurations, as introduced in 4.1 and shown in Figure 14, are examined. The results (Figure 15) indicate that sensors placed through unconstrained optimal placement provide superior reconstruction accuracy and exhibit higher stability against noise. The constrained placements achieve comparable performance to the Tall-3D configuration while requiring fewer sensors (24 vs 44), and they show even better accuracy when the same number of sensors is used. In contrast, randomly placed sensors yield pool reconstruction, characterized by both errors high and strong sensitivity to noise.

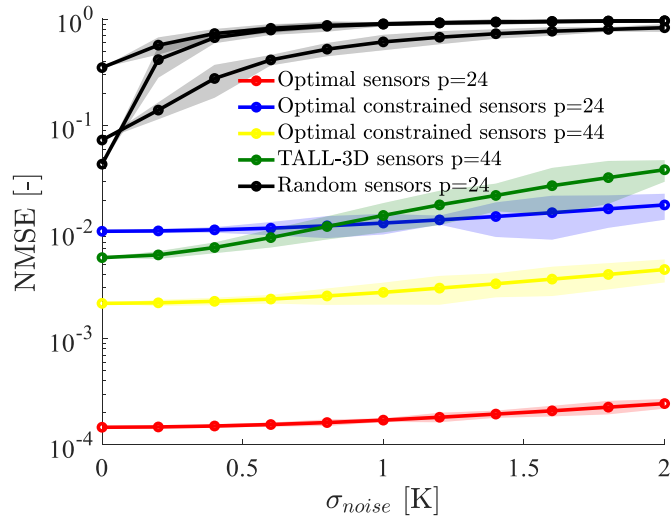

Figure 15. NMSE of reconstruction of full state temperature by sparse sensor results polluted by noise. Cases compared with different sensor layouts. Each NMSE corresponds to the mean over 10 independent reconstructions at a given noise level, with the shaded area indicating the $\pm 1\sigma$ standard deviation.

The relatively poorer reconstruction performance—approximately two orders of magnitude lower—observed for constrained sensor placement compared to the unconstrained case can be explained by examining the predicted mode coefficients $\hat{a}$ (Eqs. (3)(5)), as shown in Figure 16. In the constrained layout, sensors lack access to measurements along the central axis of the test section, which reduces the accuracy of coefficient predictions, particularly for modes 2, 3, and 6. Among these, mode 2 plays a dominant role in the data variability and represents the inertial-driven flow regime (Figure 11). Measurements taken exclusively from the outer regions of the test section are insufficient to capture this phenomenon. This analysis highlights that the effectiveness of sensor placement is strongly dependent on the specific UIPs of the sensitivity analysis that define the dataset.

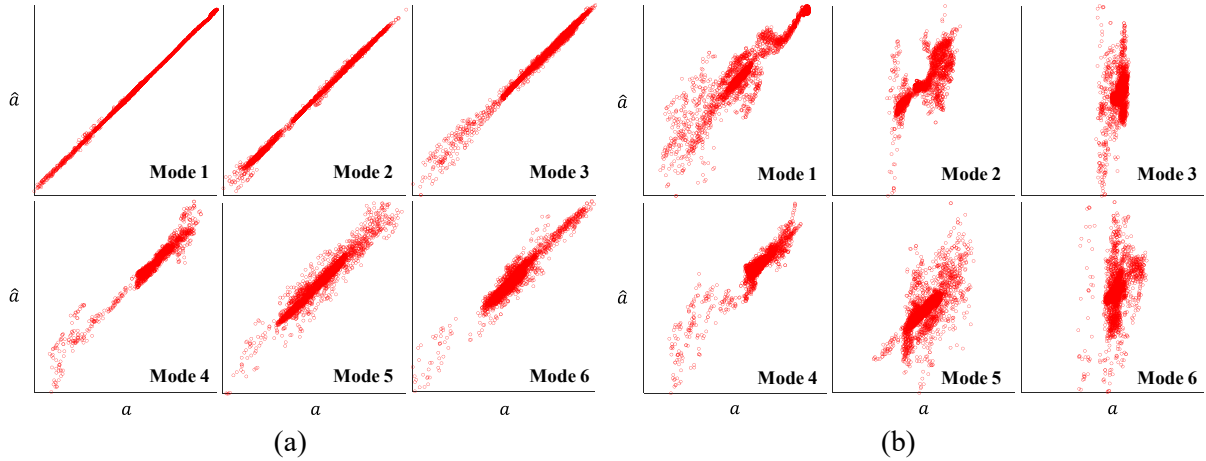

Figure 16. Comparison of mode coefficients decoded by POD ($a$) with the predicted coefficients ($\hat{a}$) by sparse sensor measurements using 24 sensors by (a) unconstrainted and (b) constrained placements.

## 5. CONCLUSIONS

The measurements in thermal-hydraulic experiments are typically obtained by sparsely distributed sensors, which provide limited coverage of the domain of interest and phenomena of interest. They are also susceptible to both systematic and random noise. The spatial placement of these sensors determines the utility of the data for model development, calibration and validation.

This study presents a data-driven methodology for optimizing sensor placement in thermal-hydraulic experiments. The proposed framework combines a sensitivity analysis to construct the dataset, POD for reduced-order modeling, and QR factorization with column pivoting to identify optimal sensor configurations while considering spatial constraints. The methodology was demonstrated on an experiment conducted in the TALL-3D LBE loop. The optimally placed TCs exhibited pronounced sensitivity to variations in UIPs and provided accurate full field reconstruction while preserving robustness in the presence of measurement noise. These findings confirm that strategically positioned sensors can significantly enhance the utility of experimental measurements for code development and validation, providing improved coverage of key thermal-hydraulic phenomena such as thermal stratification, mixing, and jet impingement.

However, unlike computer-aided engineering where sensors can be virtually positioned with minimal effort (a few clicks), practical implementation must consider installation feasibility and other engineering constraints. In practice, the most informative monitoring locations can be first identified through the proposed optimization techniques and subsequently refined through expert judgment. The rationality of such designs can be quantitatively assessed by comparing their reconstruction errors of the full-state dataset, particularly under the presence of noise.

## ACKNOWLEDGMENTS

The computations were enabled by resources provided by the National Academic Infrastructure for Supercomputing in Sweden (NAISS) and the Swedish National Infrastructure for Computing (SNIC).

# REFERENCES

1. RELAP5/MOD3.3, code manual, volume II: User's guide and input requirements, Information Systems Laboratories, Inc, Idaho, 2001.
2. GOTHIC containment analysis package, user manual, Version 8.4 (QA), EPRI, 2022.
3. ANSYS® Fluent Theory Guide, Release 2021 R2.
4. D. Paladino, R. Kapulla, S. Paranjape, S. Suter, M. Andreani. "PANDA experiments within the OECD/NEA HYMERES-2 project on containment hydrogen distribution, thermal radiation and suppression pool phenomena," Nuclear Engineering and Design, 392, 111777, (2022).
5. D. Paladino, R. Kapulla, S. Paranjape, et al. "PANDA experimental database and further needs for containment analyses," Nuclear Engineering and Design, 404, 112173, (2023).
6. C. Introini, S. Riva, S. Lorenzi, S. Cavalleri. A. Cammi. "Non-intrusive system state reconstruction from indirect measurements: A novel approach based on Hybrid Data Assimilation methods," Annals of Nuclear Energy, 182, 109538 (2023).
7. X. Wang, D, Grishchenko, P. Kudinov, et al. "Analysis of thermal stratification and erosion phenomena induced by steam injection through a sparger in large scale pool experiments PANDA and PPOOLEX," Applied Thermal Engineering, 277, 127099, (2025).
8. R. Kapulla, D. Uong, C. Zimmer, D. Paladino, "PIV measurements in the vicinity of a steam sparger in the PANDA facility," Nuclear Engineering and Design, 336, 112-121, (2018).
9. J. Laine, M. Puustinen, A. Räsänen. "PPOOLEX Experiments with a Sparger," Nordic Nuclear Safety Research, NKS-334, (2015).
10. D. Grishchenko, M. Jeltsov, K. Kööp, et al. "The TALL-3D Facility Design and Commissioning Tests for Validation of Coupled STH and CFD Codes," Nuclear Engineering and Design, 290, 144 (2015).
11. M. Jeltsov, D. Grishchenko, P. Kudinov, "Validation of Star-CCM+ for liquid metal thermal-hydraulics using TALL-3D experiment," Nuclear Engineering and Design, 341, 306-325, (2019).
12. X. Wang, D. Grishchenko, P. Kudinov, et al., "Development of Computationally Effective Models for Simulation of Steam Injection Effects on the Pool Stratification and Mixing," Available at SSRN: http://dx.doi.org/10.2139/ssrn.5051182. (2025).
13. A. Krause, A. Singh, and C. Guestrin, "Near-optimal sensor placements in Gaussian processes: Theory, efficient algorithms and empirical studies," J. Mach. Learn. Res., 9(2), 1–10, (2008).
14. S. Joshi and S. Boyd, "Sensor selection via convex optimization," IEEE Trans. Signal Process., 57(2), 451–462, (2009).
15. T. H. Summers, F. L. Cortesi, and J. Lygeros, "On submodularity and controllability in complex dynamical networks," IEEE Trans. Control Netw. Syst., 3(1), 91–101, (2016).
16. X. Wang, Y. Chan, K. Wong, D. Grishchenko, P. Kudinov. "Flow reconstruction of single-phase planar jet from sparse temperature measurements," SCOPE-1, Daharan, Saudi Arabia, November 12-15, (2023).
17. K. Manohar, B. W. Brunton, J. N. Kutz, and S. L. Brunton, "Data-driven sparse sensor placement for reconstruction: Demonstrating the benefits of exploiting known patterns," IEEE Control Syst. Mag., 38(3), 63–86, (2018).
18. N. Karnik, M. G. Abdo, C. E. Estrada-Perez, et al., "Constrained Optimization of Sensor Placement for Nuclear Digital Twins," IEEE Sensors. J. 24(9), (2024).
19. R. Everson and L. Sirovich, "Karhunen–loéve procedure for gappy data," J. Opt. Soc. Amer. A, Opt. Image Sci., vol. 12, no. 8, pp. 1657–1664, (1995).
20. Z. Drmac and S. Gugercin, "A new selection operator for the discrete empirical interpolation method: Improved a priori error bound and extensions," SIAM J. Sci. Comput., vol. 38, no. 2, pp. A631–A648, (2016).
21. P. Businger and G. H. Golub, "Linear least squares solutions by Householder transformations," Numer. Math., vol. 7, no. 3, pp. 269–276, (1965).
22. P. Kudinov, D. Grishchenko, I. Mickus, K. Kööp, M. Jeltsov. "TALL-3D setup for first series," SESAME progress report, D4.5, (2016).
23. OECD/NEA Nuclear Science Committee, 2015. Handbook on Lead-bismuth Eutectic Alloy and Lead Properties, Materials Compatibility, Thermal-hydraulics and Technologies.

24. Morris, M.D., "Factorial Sampling Plans for Preliminary Computational Experiments," Technometrics 33, 161–174, (1991).
25. Adams, T., et al., "Multilevel Parallel Object-Oriented Framework for Design Optimization, Parameter Estimation, Uncertainty Quantification, and Sensitivity Analysis: Version 6.0 User's Manual," Sandia Technical Report SAND2014-4633 (2009).